\documentclass[twocolumn,prl,showpacs]{revtex4}

\usepackage{graphicx}
\usepackage{amsmath}

\begin{document}

\newcommand{\be}{\begin{equation}}
\newcommand{\ee}{\end{equation}}
\newcommand{\bfig}{\begin{figure}}
\newcommand{\efig}{\end{figure}}
\newcommand{\bea}{\begin{eqnarray}}
\newcommand{\eea}{\end{eqnarray}}
\newcommand{\infinitessimal}{\mathrm{d}}
\newcommand{\infinitesmal}{\mathrm{d}}
\newcommand{\infinitesimal}{\mathrm{d}}
\newcommand{\intd}{\mathrm{d}}

\newcommand{\raa}{$R_{AA}$ }
\newcommand{\raacomma}{$R_{AA}$}
\newcommand{\raaphi}{$R_{AA}(\phi)$ }
\newcommand{\raaphicomma}{$R_{AA}(\phi)$}
\newcommand{\raaphipt}{$R_{AA}(\phi;\,\eqnpt)$ }
\newcommand{\raaphiptcomma}{$R_{AA}(\phi;\,\eqnpt)$} 
\newcommand{\raapt}{$R_{AA}(\eqnpt)$ } 
\newcommand{\raaptcomma}{$R_{AA}(\eqnpt)$} 
\newcommand{\eqnraapt}{R_{AA}(\eqnpt)} 
\newcommand{\raaq}{$R_{AA}^q$ }
\newcommand{\raaqcomma}{$R_{AA}^q$}
\newcommand{\eqnraaq}{R_{AA}^q}
\newcommand{\raaqphi}{$R_{AA}^q(\phi)$ }
\newcommand{\raaqphicomma}{$R_{AA}^q(\phi)$} 
\newcommand{\eqnraaqphi}{R_{AA}^q(\phi)}
\newcommand{\raaqphipt}{$R_{AA}^q(\phi;\,\eqnpt)$ }
\newcommand{\raaqphiptcomma}{$R_{AA}^q(\phi;\,\eqnpt)$} 
\newcommand{\eqnraaqphipt}{R_{AA}^q(\phi;\,\eqnpt)} 
\newcommand{\raaqpt}{$R_{AA}^q(\eqnpt)$ } 
\newcommand{\raaqptcomma}{$R_{AA}^q(\eqnpt)$} 
\newcommand{\eqnraaqpt}{R_{AA}^q(\eqnpt)}

\newcommand{\raac}{$R_{AA}^c$ }
\newcommand{\raaccomma}{$R_{AA}^c$}
\newcommand{\eqnraac}{R_{AA}^c}
\newcommand{\raacphi}{$R_{AA}^c(\phi)$ }
\newcommand{\raacphicomma}{$R_{AA}^c(\phi)$} 
\newcommand{\eqnraacphi}{R_{AA}^c(\phi)}
\newcommand{\raacphipt}{$R_{AA}^c(\phi;\,\eqnpt)$ }
\newcommand{\raacphiptcomma}{$R_{AA}^c(\phi;\,\eqnpt)$} 
\newcommand{\eqnraacphipt}{R_{AA}^c(\phi;\,\eqnpt)} 
\newcommand{\raacpt}{$R_{AA}^c(\eqnpt)$ } 
\newcommand{\raacptcomma}{$R_{AA}^c(\eqnpt)$} 
\newcommand{\eqnraacpt}{R_{AA}^c(\eqnpt)} 

\newcommand{\raab}{$R_{AA}^b$ }
\newcommand{\raabcomma}{$R_{AA}^b$}
\newcommand{\eqnraab}{R_{AA}^b}
\newcommand{\raabphi}{$R_{AA}^b(\phi)$ }
\newcommand{\raabphicomma}{$R_{AA}^b(\phi)$} 
\newcommand{\eqnraabphi}{R_{AA}^b(\phi)}
\newcommand{\raabphipt}{$R_{AA}^b(\phi;\,\eqnpt)$ }
\newcommand{\raabphiptcomma}{$R_{AA}^b(\phi;\,\eqnpt)$} 
\newcommand{\eqnraabphipt}{R_{AA}^b(\phi;\,\eqnpt)} 
\newcommand{\raabpt}{$R_{AA}^b(\eqnpt)$ } 
\newcommand{\raabptcomma}{$R_{AA}^b(\eqnpt)$} 
\newcommand{\eqnraabpt}{R_{AA}^b(\eqnpt)}

\newcommand{\cbratio}{$\eqnraacpt/\eqnraabpt$ }
\newcommand{\cbratiocomma}{$\eqnraacpt/\eqnraabpt$}
\newcommand{\eqncbratio}{\eqnraacpt/\eqnraabpt}

\newcommand{\raag}{$R_{AA}^g$ }
\newcommand{\raagcomma}{$R_{AA}^g$}
\newcommand{\eqnraag}{R_{AA}^g}
\newcommand{\raagphi}{$R_{AA}^g(\phi)$ }
\newcommand{\raagphicomma}{$R_{AA}^g(\phi)$} 
\newcommand{\eqnraagphi}{R_{AA}^g(\phi)}
\newcommand{\raagphipt}{$R_{AA}^g(\phi;\,\eqnpt)$ }
\newcommand{\raagphiptcomma}{$R_{AA}^g(\phi;\,\eqnpt)$} 

\newcommand{\eqnraagphipt}{R_{AA}^g(\phi;\,\eqnpt)} 
\newcommand{\raagpt}{$R_{AA}^g(\eqnpt)$ } 
\newcommand{\raagptcomma}{$R_{AA}^g(\eqnpt)$} 
\newcommand{\eqnraagpt}{R_{AA}^g(\eqnpt)} 

\newcommand{\RAA}{\raa}
\newcommand{\RAAcomma}{\raacomma}
\newcommand{\RAAphi}{\raaphi}
\newcommand{\RAAphicomma}{\raaphicomma}
\newcommand{\RAAphipt}{\raaphipt}
\newcommand{\RAAphiptcomma}{\raaphiptcomma}
\newcommand{\raapi}{$R_{AA}^\pi$ }
\newcommand{\raae}{$R_{AA}^{e^-}$ }
\newcommand{\raapicomma}{$R_{AA}^\pi$}
\newcommand{\raaecomma}{$R_{AA}^{e^-}$}
\newcommand{\eqnraapi}{R_{AA}^\pi}
\newcommand{\eqnraae}{R_{AA}^{e^-}}

\newcommand{\vtwo}{$v_2$ }
\newcommand{\vtwocomma}{$v_2$}
\newcommand{\vtwopt}{$v_2(\eqnpt)$ }
\newcommand{\vtwoptcomma}{$v_2(\eqnpt)$}
\newcommand{\eqnraa}{R_{AA}}
\newcommand{\eqnraaphi}{R_{AA}(\phi)}
\newcommand{\eqnraaphipt}{R_{AA}(\phi;\,\eqnpt)} 
\newcommand{\eqnRAA}{\eqnraa}
\newcommand{\eqnRAAphi}{\eqnraaphi}
\newcommand{\eqnRAAphipt}{\eqnraaphipt}
\newcommand{\eqnvtwo}{v_2}
\newcommand{\vtwovsraa}{\vtwo vs.~\raa}
\newcommand{\vtwovsraacomma}{\vtwo vs.~\raacomma}
\newcommand{\eqnvtwopt}{v_2(\eqnpt)}

\newcommand{\pp}{$p+p$ }
\newcommand{\ppcomma}{$p+p$}
\newcommand{\dau}{$d+Au$ }
\newcommand{\daucomma}{$d+Au$}
\newcommand{\auau}{$Au+Au$ }
\newcommand{\auaucomma}{$Au+Au$}
\newcommand{\aplusa}{$A+A$ }
\newcommand{\aplusacomma}{$A+A$}
\newcommand{\cucu}{$Cu+Cu$ }
\newcommand{\cucucomma}{$Cu+Cu$}

\newcommand{\rhopart}{$\rho_{\textrm{\footnotesize{part}}}$ }
\newcommand{\rhopartcomma}{$\rho_{\textrm{\footnotesize{part}}}$}
\newcommand{\eqnrhopart}{\rho_{\textrm{\footnotesize{part}}}}
\newcommand{\npart}{$N_{\textrm{\footnotesize{part}}}$ }
\newcommand{\npartcomma}{$N__{\textrm{\footnotesize{part}}}$}
\newcommand{\eqnnpart}{N_{\textrm{\footnotesize{part}}}}
\newcommand{\taa}{$T_{AA}$ }
\newcommand{\taacomma}{$T_{AA}$}
\newcommand{\eqntaa}{T_{AA}}
\newcommand{\rhocoll}{$\rho_{\textrm{\footnotesize{coll}}}$ }
\newcommand{\rhocollcomma}{$\rho_{\textrm{\footnotesize{coll}}}$}
\newcommand{\eqnrhocoll}{\rho_{\textrm{\footnotesize{coll}}}}
\newcommand{\ncoll}{$N_{\textrm{\footnotesize{coll}}}$ }
\newcommand{\ncollcomma}{$N_{\textrm{\footnotesize{coll}}}$}
\newcommand{\eqnncoll}{N_{\textrm{\footnotesize{coll}}}}
\newcommand{\dndy}{$\frac{dN_g}{dy}$ }
\newcommand{\dndycomma}{$\frac{dN_g}{dy}$}
\newcommand{\eqndndy}{\frac{dN_g}{dy}}
\newcommand{\eqndndyabs}{\frac{dN_g^{abs}}{dy}}
\newcommand{\eqndndyrad}{\frac{dN_g^{rad}}{dy}}
\newcommand{\dnslashdy}{$dN_g/dy$ }
\newcommand{\dnslashdycomma}{$dN_g/dy$}
\newcommand{\eqndnslashdy}{dN_g/dy}
\newcommand{\as}{\alpha_s}
\newcommand{\alphas}{$\as$ }
\newcommand{\alphascomma}{$\as$}
\newcommand{\eqnalphas}{\as}

\newcommand{\pt}{$p_T$ }
\newcommand{\pT}{\pt}
\newcommand{\ptcomma}{$p_T$}
\newcommand{\pTcomma}{\ptcomma}
\newcommand{\eqnpt}{p_T}
\newcommand{\ptf}{$p_T^f$ }
\newcommand{\ptfcomma}{$p_T^f$}
\newcommand{\eqnptf}{p_T^f}
\newcommand{\pti}{$p_T^i$ }
\newcommand{\pticomma}{$p_T^i$}
\newcommand{\eqnpti}{p_T^i}
\newcommand{\lowpt}{low-\pt}
\newcommand{\lowptcomma}{low-\ptcomma}
\newcommand{\midpt}{mid-\pt}
\newcommand{\midptcomma}{mid-\ptcomma}
\newcommand{\intermediatept}{intermediate-\pt}
\newcommand{\intermediateptcomma}{intermediate-\ptcomma}
\newcommand{\highpt}{high-\pt}
\newcommand{\highptcomma}{high-\ptcomma}
\newcommand{\Aperp}{$A_\perp$ }
\newcommand{\Aperpcomma}{$A_\perp$}
\newcommand{\eqnAperp}{A_\perp}
\newcommand{\rperp}{$r_\perp$ }
\newcommand{\rperpcomma}{$r_\perp$}
\newcommand{\eqnrperp}{r_\perp}
\newcommand{\eqnrperpHS}{r_{\perp,HS}}
\newcommand{\eqnrperpWS}{r_{\perp,WS}}
\newcommand{\Rperp}{$R_\perp$ }
\newcommand{\Rperpcomma}{$R_\perp$}
\newcommand{\eqnRperp}{R_\perp}

\newcommand{\pizero}{$\pi^0$ }
\newcommand{\eqnpizero}{\pi^0}

\newcommand{\qhat}{$\hat{q}$ }
\newcommand{\qhatcomma}{$\hat{q}$}
\newcommand{\eqnqhat}{\hat{q}}

\newcommand{\gym}{$g_{SYM}$ }
\newcommand{\gymcomma}{$g_{SYM}$}
\newcommand{\eqngym}{g_{SYM}}
\newcommand{\gsym}{\gym}
\newcommand{\gsymcomma}{\gymcomma}
\newcommand{\eqngsym}{\eqngym}
\newcommand{\gs}{$g_{s}$ }
\newcommand{\gscomma}{$g_{s}$}
\newcommand{\eqngs}{g_{s}}
\newcommand{\ads}{AdS/CFT }
\newcommand{\asd}{\ads}
\newcommand{\adscomma}{AdS/CFT}
\newcommand{\asdcomma}{\adscomma}
\newcommand{\asym}{$\alpha_{SYM}$ }
\newcommand{\asymcomma}{$\alpha_{SYM}$}
\newcommand{\eqnasym}{\alpha_{SYM}}
\newcommand{\alphasym}{\asym}
\newcommand{\alphasymcomma}{\asymcomma}
\newcommand{\eqnalphasym}{\eqnalsym}

\newcommand{\infinity}{\infty}

\newcommand{\eq}[1]{Eq.~(\ref{#1})}
\newcommand{\eqn}[1]{Eq.~(\ref{#1})}
\newcommand{\fig}[1]{Fig.~\ref{#1}}
\newcommand{\figtwo}[2]{Figs.~\ref{#1}, \ref{#2}}
\newcommand{\tab}[1]{Table \ref{#1}}
\newcommand{\captionsize}{\small}

\title{Testing 
 \ads Deviations from pQCD Heavy Quark Energy Loss \\ with
Pb+Pb at LHC}

\date{\today}

\author{W. A. Horowitz}
\email{horowitz@phys.columbia.edu}
\affiliation{Frankfurt Institute for Advanced Studies (FIAS), 60438 Frankfurt am Main, Germany}
\affiliation{Department of Physics, Columbia University,
             538 West 120$^{th}$ Street, New York, NY 10027, USA}
\author{M. Gyulassy}
\email{gyulassy@fias.uni-frankfurt.de}
\affiliation{Frankfurt Institute for Advanced Studies (FIAS), 60438 Frankfurt am Main, Germany}
\affiliation{Department of Physics, Columbia University,
             538 West 120$^{th}$ Street, New York, NY 10027, USA}

\begin{abstract}
  Heavy quark jet quenching in nuclear collisions at LHC 
is predicted and compared using the classical gravity \ads
correspondence and Standard Model  
  perturbative QCD.  
The momentum independence and inverse quark mass dependence
  of the drag coefficient in \ads differs substantially from the 
characteristic
  $\log(\eqnpt/M)/\eqnpt$ variation of the drag in QCD.  We propose
  that the measurement of the momentum dependence of the double ratio
  of the nuclear modification factors of charm and bottom jets,
  \cbratiocomma, is a robust observable that can be used to search for strong coupling deviations from perturbative QCD predictions.
\end{abstract}
\pacs{12.38.Mh; 24.85.+p; 25.75.-q}

\maketitle
{\em Introduction.} Recent discoveries
 at RHIC \cite{Adcox:2004mh,GMc,Riordan:2006df}
have led to provocative suggestions \cite{Kovtun:2004de} that
the properties of strongly coupled 
quark gluon plasmas (sQGP) produced in ultra-relativistic
nuclear collisions may be better approximated by
string theoretic inspired AdS/CFT gravity-gauge theory
correspondences \cite{Maldacena:1997re} 
than conventional Standard Model perturbative QCD (pQCD). 
Four  main classes of observables have attacted
the most attention: (1) Entropy production
as probed by multiplicity distributions \cite{Back:2001ae}, (2) Perfect Fluidity \cite{Riordan:2006df} as 
probed by collective elliptic flow
observables \cite{STARv2,Huovinen:2001cy,Molnar:2001ux}, 
(3) Jet Quenching as probed by \highpt
hadrons \cite{PhenixJets,Wang:1991xy} and leptons
\cite{Abelev:2006db,Adare:2006nq}, 
and (4) Dijet-Bulk Correlations
as probed by two and three particle correlations \cite{Adler:2002tq,
Adler:2005ee}. 

Qualitative  successes of recent \ads applications \cite{Kovtun:2004de}
to nuclear collision phenomenology 
include the analytic account for (1) 
the surprisingly small ($\sim$ 3/4) drop \cite{GubserEntrop}
 of the entropy density in lattice QCD calculations
relative to Stefan-Boltzmann, 
(2) the order of magnitude reduction of the viscosity to entropy ratio
$\eta/s$ predicted relative to pQCD needed
to explain the seemingly near perfect
fluid flow of the sQGP observed at RHIC, (3) the 
surprising large stopping power of high transverse momenta
heavy
quarks as inferred from the 
quenching and elliptic flow of leptonic decay fragments
of heavy quark jets, and (4) the possible occurrence of 
conical ``Mach'' wave-like correlations
of hadrons associated with jets.

While quantitative and systematic comparisons of \ads gravity dual models
with  nuclear collision data are still incomplete  and 
while the conjectured
{\em double} %
Type IIB string theory$\leftrightarrow$conformal
 Supersymmetric Yang-Mills (SYM) gauge theory$\leftrightarrow$non-conformal, non-supersymmetric QCD correspondence
remains under debate (see,
e.g. \cite{McLerran:2007hz}),
the current successes provide strong  motivation
to seek more sensitive experimental tests 
that could help guide the development
of such novel theoretical approaches.
 
The aim of this Letter is to propose a robust new test that could
reveal possible strong coupling deviations from pQCD 
predictions as suggested by a specific version
of the \ads correspondence. This test involves the nuclear modification
of  identified heavy quark jets
produced in central Pb+Pb reactions at 5.5 ATeV at LHC.  Similar tests can be performed at RHIC, after future detector upgrades, and will be reported elsewhere. Specifically, we
propose that the double ratio of identified charm and bottom jet
nuclear modification factors $R_{AA}^Q(\eqnpt)$ can 
easily distinguish between a wide class of pQCD models and 
a class of gravity
dual models \cite{Karch02,Herzog:2006gh,Gubser:2006bz,
Gubser:2006nz,Gubser:2007nd} of heavy quark dynamics.  The current failure of
pQCD models to account \underline{\em quantitatively} for the recent RHIC data from STAR
\cite{Abelev:2006db} and PHENIX \cite{Adare:2006nq} on the
nonphotonic electron spectrum
provides additional strong motivation to focus on heavy quark jet observables.
Unlike for 
light quark and gluon jet observables, where pQCD predictions
were found to be remarkably quantitative
\cite{Wang:1991xy}, 
heavy quark jet quenching, especially as inferred indirectly for bottom quarks,
appears to be significantly underpredicted \cite{Dokshitzer:2001zm,Djordjevic:2005db,
Wicks:2005gt,Armesto:2005mz}. However, the pQCD predictions can not
yet be falsified at RHIC because of (1) remaining
uncertainty in
the nuclear production ratio of bottom to charm
quarks and (2) current controversy
over the relative magnitude of elastic versus radiative loss channels
\cite{Wicks:2005gt,Armesto:2005mz}.
Recent attempts to reconcile pQCD predictions
with the data by adding nonperturbative hadronic
final state interaction effects 
can be found in \cite{Adil:2006ra,vanHees:2005wb}.

The AdS/CFT correspondence has so far been applied to QCD jet physics 
in three different
ways. The first calculates the Wilson line correlator needed to
predict the partonic radiative transport coefficient \qhat \cite{Liu:2006ug,
Armesto:2003jh}. The second predicts
the heavy quark diffusion
coefficient $D$ \cite{Casalderrey-Solana:2006rq} used as an input
parameter in a relativistic Langevin model with drag 
deduced from the Einstein relation
\cite{Moore:2004tg}. The third 
computes directly the heavy quark drag
coefficient via a specific classical string configuration \cite{Herzog:2006gh,Gubser:2006bz,Gubser:2006nz}.  
We note that
the first two approaches conform more closely to the original
 spirit of the AdS/CFT conjecture relating quantum SYM correlation functions
to the asymptotic behavior of the classical supergravity (SUGRA) correlations. 
Unfortunately these correspondences are hard
to interpret in terms of gauge theory energy loss mechanisms since
 infinitely coupled SYM does not support a familiar
quasiparticle basis similar to   
gluon and quark degrees of freedom in QCD. 
In contrast, the third model, while more easily interpretable, requires a stronger form of the \ads correspondence. 
All three approaches remain under active debate (see, e.g. \cite{Gubser:2006nz,Liu:2006he,Bertoldi:2007sf}).

We focus in this Letter on the third proposed AdS/
CFT application
that involves the most direct 
string theoretic inspired gravity ``realization'' 
of heavy quark dynamics \cite{Karch02,Herzog:2006gh,Gubser:2006bz}.
A heavy quark in the fundamental representation
is a bent Nambu-Goto string with 
one end attached to a probe brane 
and that trails back above the horizon
of a D3 black brane representing the 
uniform strongly coupled SYM plasma heat bath. 
This geometry
maps the drag force problem
into a modern string theoretic  
version of the old 1696 Brachistochrone
problem that yields a remarkable, simple analytic solution for the string shape and momentum loss
per unit time.

\emph{AdS/CFT compared to pQCD} 
Exploiting this AdS/
CFT correspondence, the drag coefficient for
a massive quark moving through a strongly-coupled SYM plasma in the $\lambda =\eqngym^2N_c\gg 1$, $N_c\gg 1$, $M_Q\gg T^*$
limit is given in \cite{Herzog:2006gh,Gubser:2006bz,Gubser:2006nz} 
as  
\be
\label{mu}
\frac{d\eqnpt}{dt} = -\mu_Q \eqnpt = -
\frac{\pi\sqrt{\lambda}(T^*)^2}{2M_Q}\eqnpt, \ee
where $T^*$ is the temperature of the SYM plasma as fixed by the Hawking temperature of the dual D3 black brane.  Issues related to the relaxation of the strong assumptions made in deriving, and the momentum limitation of the applicability of, \eq{mu} will be discussed later in the text.
\bfig[!ht]
\includegraphics[width=\columnwidth]{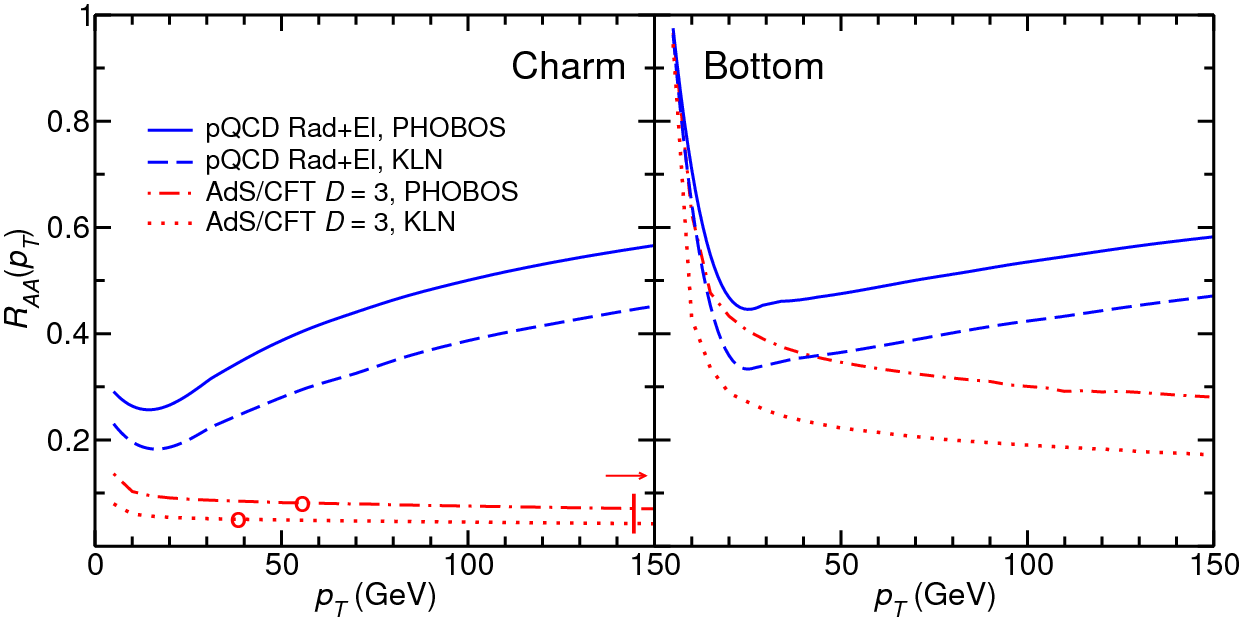}
\caption{\label{LHCcandbRAA}\captionsize{(Color Online)
\raacpt and \raabpt predicted for central Pb+Pb at LHC comparing \ads
\eq{mu} and pQCD using the WHDG model \cite{Wicks:2005gt}
convolving elastic and inelastic parton energy loss.  Possible initial gluon rapidity densities at LHC 
are given by $dN_g/dy=1750$, from a PHOBOS \cite{Back:2001ae,Adcox:2000sp} extrapolation, or $dN_g/dy=2900$, from the KLN model of the color glass condensate (CGC) \cite{Kharzeev:2004if}.
The top two curves from pQCD increase with \pt while the bottom two curves from \ads slowly decrease with \ptcomma.  The \ads parameters here were found using the ``obvious'' prescription 
with $\eqnasym=.05$, $\tau_0=1$ fm/c, giving $D/2\pi T=3$ (abbreviated to $D=3$ in the figure). 
Similar trends were seen for the other input parameter possibilities discussed
in the text.
}}  \efig

Applying \eq{mu} to LHC requires an additional proposal
that maps QCD temperatures and couplings to the SYM world
and its SUGRA dual.  The ``obvious''  first prescription \cite{Gubser:2006qh}
is to take
$\eqngym=\eqngs$ constant, $T^{*}=T^{QCD}$, and
$N_c=3$.
However it was suggested in \cite{Gubser:2006qh} that
a more physical ``alternative'' might be to equate energy
densities, giving $T^{*}=T^{QCD}/3^{1/4}$, and
to fit the coupling $\lambda=\eqngym^2N_c\approx5.5$ in order to reproduce
the static quark-antiquark forces calculated via lattice QCD.

The string theoretic result for the diffusion coefficient
used in the Langevin model is $D/2\pi T^*=4/\sqrt{\lambda}$ \cite{Casalderrey-Solana:2006rq}.  This illustrates well the problem of connecting the $T^*$ and $\lambda$ of SYM
to ``our'' QCD world. Using the
``obvious'' prescription with $\eqnalphas=.3$, $N_c=3$, one finds
$D/2\pi T\sim1.2$.  However, $D/2\pi T=3$ was claimed in
\cite{Adare:2006nq,Casalderrey-Solana:2006rq} to fit PHENIX data somewhat better. Note that
$D/2\pi T=3$
requires an unnaturally small $\eqnalphas\sim 0.05$ that is very far 
from the assumed $\lambda \gg 1$ 't Hooft limit. 

We proceed by computing the nuclear modification factors,
neglecting initial state shadowing or saturation effects.
In order to correctly 
deconvolute such effects from the final state effects
that we compute below, it will be necessary to measure nuclear
modification factors in $p+A$ as a function of 
$(y,\eqnpt)$ at LHC just as $d+A$ was the critical
control experiment \cite{Adcox:2004mh} at RHIC \cite{GMc}.

Final state suppression of \highpt jets due to a fractional energy loss
$\epsilon$, $\eqnptf=(1-\epsilon)\eqnpti$, can be computed knowing 
the $Q$-flavor dependent spectral indices
$n_Q+1=-\frac{d}{d\log \eqnpt}\log\left(\frac{d\sigma_Q}{dyd\eqnpt}\right)$
from pQCD or directly from $p+p\rightarrow Q+X$ data. 
The nuclear modification factor is then
$R_{AA}^Q(\eqnpt)=<(1-\epsilon)^{n_Q}>$, where the average
is over the distribution $P(\epsilon;M_Q,\eqnpt,\ell)$ that depends
in general on the quark mass, \ptcomma, and the path length $\ell$ of the
jet through the sQGP. As in \cite{Wicks:2005gt}
we average over jets produced according to the binary distribution
geometry and compute $\ell$ through a participant
transverse density distribution taking into account the nuclear diffuseness.
Given $dN_g/dy$ of produced gluons, the temperature is computed
assuming isentropic Bjorken $1D$ Hubble flow.
As emphasized in \cite{Wicks:2005gt}, detailed geometric path length averaging
plays a crucial role in allowing consistency between $\pi^0,\eta$ and heavy quark
quenching in pQCD.

For \ads drag, \eq{mu} gives the average fractional
energy loss as $\bar{\epsilon}=1-\exp(-\mu_Q\ell)$.  Energy loss 
is assumed to start at thermalization, $\tau_0\sim0.6-1.0$ fm/c, and stops when the
confinement  temperature, $T_c\sim160$ MeV, is reached.  The exponentiated $T^2$ dependence in $\mu_Q$ leads to a significant sensitivity to the opacity of the medium, as well as to $\tau_0$ and $T_c$.

To understand the generic qualitative 
features of our numerical results
it is instructive to consider the simplest case of a geometric
path average over a static, finite, uniform plasma of thickness $L$; then
\be
R_{AA}^Q(\eqnpt)=\frac{1-e^{n_Q\mu_Q L}}{n_Q \mu_Q L}\approx\frac{1}{n_Q \mu_Q L},
\label{aprx}
\ee
where the \pt dependence is carried entirely by the spectral index $n_Q(\eqnpt)$.
\raa can be interpreted for $L\gg \ell_Q\equiv 1/(n_Q\mu_Q)$
 as the fraction 
$\ell_Q/L$ of the Q jets that escape
unstopped from the strongly coupled plasma 
within the \ads approximation.

\bfig[!ht]
\includegraphics[width=\columnwidth]{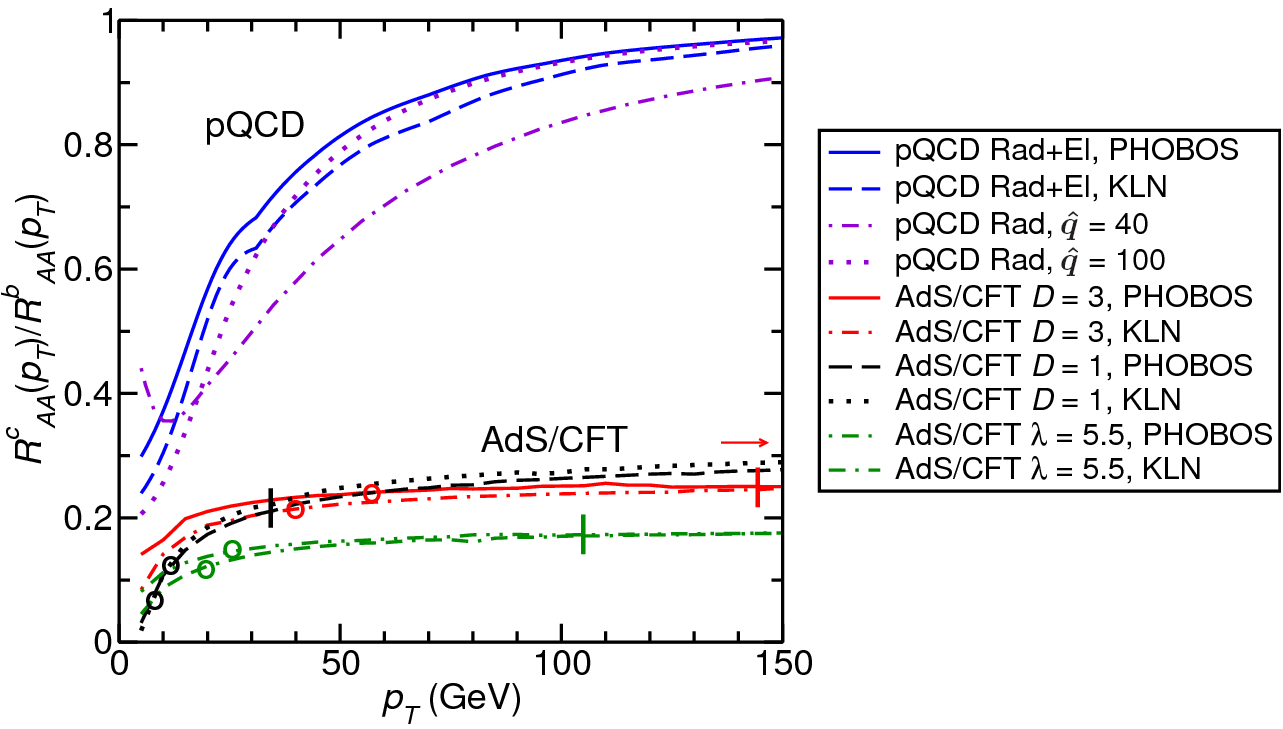}
\caption{\label{Ratio}\captionsize{The double ratio
of \raacpt to \raabpt predictions for LHC using \eq{mu} for
\ads and WHDG \cite{Wicks:2005gt} for pQCD with a wide range
of input parameters. The generic difference between the pQCD results
tending to unity contrasted to
the much smaller and nearly \ptcomma-independent results from \ads
can be easily distinguished at LHC.}}
\efig

Two implementations of pQCD energy loss are used in this paper.  The
first is the full WHDG model convolving fluctuating
 elastic and inelastic loss with fluctuating path geometry
\cite{Wicks:2005gt}.  The second restricts WHDG to include
only radiative loss in order to facilitate comparison to 
\cite{Armesto:2003jh}.
Note that 
when
realistic nuclear geometries with Bjorken expansion are used, the
``fragility'' of \raa for large \qhat reported in \cite{Eskola:2004cr}
is absent in both implementations of WHDG.

Unlike the \ads dynamics, pQCD predicts \cite{Dokshitzer:2001zm,
Djordjevic:2005db,Wicks:2005gt} 
that the average energy loss fraction
in a static uniform plasma is approximately
$\bar{\epsilon} \approx 
\kappa L^2 \hat{q}
 \log(\eqnpt/M_Q)/\eqnpt$, 
with $\kappa$ a proportionality constant and $\hat{q}=\mu_D^2/\lambda_g$.
The most important feature in pQCD relative to \ads is that
$\bar{\epsilon}_{pQCD}\rightarrow 0$ asymptotically at \highpt while
$\bar{\epsilon}_{AdS}$ remains constant.  $n_Q(\eqnpt)$
is a slowly increasing function of momentum; thus $R_{AA}^{pQCD}$
increases with \pt whereas $R_{AA}^{AdS}$ decreases.  This generic difference can be observed in
\fig{LHCcandbRAA}, which shows representative predictions from the full
numerical calculations of charm and bottom \raapt at LHC.

\emph{Double Ratio of charm to bottom $R_{AA}^Q$} 
A disadvantage of the $R_{AA}^Q(\eqnpt)$ observable alone
is that its normalization and slow 
\pt dependence can be fit with different model assumptions
compensated by using very different medium parameters. 
In particular, high value extrapolations
of the \qhat parameter
proposed in \cite{Armesto:2005mz} could simulate
the flat \pt independent prediction from AdS/CFT. 

We propose to use the double ratio of charm to bottom \raa to amplify
the observable difference between the mass and \pt dependencies of the
\ads drag and pQCD-inspired energy loss models.  One can see in
\fig{Ratio} that not only are most overall normalization differences
canceled, but also that the curves remarkably bunch to either AdS/CFT-like 
or pQCD-like generic
results regardless of the input parameters used.

The numerical value of $R^{cb}$ shown in \fig{Ratio} for \ads
 can be roughly understood analytically from
\eq{aprx}
as, 
\be
R^{cb}_{AdS} 
\approx \frac{M_c}{M_b} \frac{n^b(\eqnpt)}{n^c(\eqnpt)}
\approx\frac{M_c}{M_b}\approx 0.26 \label{cbads},
\ee
where in this approximation all $\lambda$, $T^*$, $L$, and 
$n_c(\eqnpt)\approx n_b(\eqnpt)$ dependences drop out. 

The pQCD trend in \fig{Ratio} can be understood qualitatively
from the expected behavior of
$\bar{\epsilon}_{pQCD}$ noted above
giving (with $n_c\approx n_b=n$)
\be
\label{pQCDratio}
R^{cb}_{pQCD} \approx 1 - \frac{p_{cb}}{\eqnpt},
\ee
where $p_{cb}=\kappa
n(\eqnpt) L^2 \log(M_b/M_c) \eqnqhat$ sets the relevant
momentum scale.  Thus $R^{cb}\rightarrow1$ more slowly for higher opacity.
One can see this behavior reflected in the full numerical results shown in \fig{Ratio} for moderate suppression, but that the extreme opacity
 $\eqnqhat=100$ case deviates from \eq{pQCDratio}.
 
 The maximum momentum for which string theoretic predictions for
 $R^{cb}$ can be trusted is not well understood.  \eq{mu} was derived
 assuming a constant heavy quark velocity.  Supposing this is
 maintained by the presence of an electromagnetic field, the
 Born-Infeld action gives a ``speed limit'' of $\gamma_c=M^2/\lambda (T^*)^2$
 \cite{Casalderrey-Solana:2007qw}.  The work of \cite{Herzog:2006gh}
 relaxed the assumptions of infinite quark mass and constant velocity;
 nevertheless \eq{mu} well approximates the full results.  Requiring a
 time-like endpoint on the probe brane for a constant velocity string
 representing a finite mass quark leads to \cite{Gubser:2006nz} a
 parametrically similar cutoff, \be\label{speedlimit}
 \gamma_c=\left(1+\frac{2M}{\sqrt{\lambda}T^*}\right)^2\approx\frac{4M^2}{\lambda(T^*)^{2}}.
 \ee 
 There is no known limit yet for the dynamic velocity case.
To get a sense of
 the \pt scale where the \ads approximation may break down, we plot
 the momentum cutoffs from \eq{speedlimit} for the given SYM input parameters corresponding
 to $T^*(\tau_0)$ and $T^*_c$.  These are depicted by ``O'' and
 ``$|$'' in the figures, respectively.

\emph{Conclusions} Possible strong coupling deviations
from pQCD in nuclear collisions 
were studied based on a recent \ads 
 model of 
charm and bottom energy loss.  The predicted nuclear modification
factors, $R_{AA}^Q$, were
found to decrease as a function of \ptcomma, as compared to  increasing as
predicted from pQCD.  The distinction between these dynamical
models is amplified by studying the \pt dependence of the
double ratio \cbratiocomma, which clearly illustrates a quark mass independence
of pQCD energy loss at asymptotically high momenta in contrast to the
momentum independence and inverse mass dependence prediction
of the \ads drag coefficient.   
A clear signal of novel nonperturbative
physics would be the observation 
of a large deviation of the double ratio from unity
at $\eqnpt\gg M_b$ soon to be accessible at LHC and RHIC.

\section{Acknowledgments:} We are grateful to
Simon Wicks for extensive discussions.
We also thank A.~Adil, J.~Casalderry-Solana, B.~Cole, W.~Greiner, J.~Harris, H.~St\"ocker, and U.~Wiedemann
for useful discussions. 
MG acknowledges support from
the DFG, 
FIAS and ITP 
of the J.W. Goethe Uni. Frankfurt, and from 
GSI. 
 This work was also supported by the Director, Office of Science, 
Office of High Energy and Nuclear Physics, Division of Nuclear
Physics, of the U.S. Department of Energy under Grants No. DE-FG02-93ER40764.


\end{document}